\documentclass[aps,preprintnumbers,eqsecnum,amsmath,amssymb,showpacs,nofootinbib]{revtex4}
\usepackage{graphicx}
\usepackage{dcolumn
}\usepackage{bm}
\usepackage{epsfig}
\usepackage{color}

\begin{document}
\title{Annihilation type rare radiative $B_{(s)}\to V\gamma$ decays}
\author{Anastasiya Kozachuk$^a$, Dmitri Melikhov$^{a,b}$, Nikolai Nikitin$^{a,c,d}$}
\affiliation{$^a$D.~V.~Skobeltsyn Institute of Nuclear Physics, M.~V.~Lomonosov
Moscow State University, 119991, Moscow, Russia
\\ $^b$Faculty of Physics, University of Vienna, Boltzmanngasse 5, A-1090 Vienna, Austria
\\ $^c$M.~V.~Lomonosov Moscow State University, Physical Facutly, 119991, Moscow, Russia
\\ $^d$A.~I.~Alikhanov Institute for Theoretical and Experimental Physics, 117218 Moscow, Russia}
\date{\today} 
\begin{abstract}
We obtain predictions for a number of radiative decays $B_{(s)}\to V\gamma$, $V$ the vector meson, which proceed 
through the weak-annihilation mechanism. Within the factorization approximation, we take into account the photon 
emission from the $B$-meson loop and from the vector-meson loop; the latter 
subprocesses were not considered in the previous analyses but are found to have sizeable impact on the 
$B_{(s)}\to V\gamma$ decay rate. The highest branching ratios for the weak-annihilation reactions reported here are 
${\cal B}(\bar B^0_s\to J/\psi\gamma)=1.5\cdot 10^{-7}$ and  
${\cal B}(B^-\to \bar D_s^{*-}\gamma)=1.7\cdot 10^{-7}$, 
the estimated accuracy of these predictions being at the level of 20\%. 
\end{abstract}
\maketitle

\section{Introduction}
The investigation of rare $B$ decays forbidden at the tree level in the Standard Model provides the 
possibility to probe the electroweak sector at large mass scales. Interesting information 
about the structure of the theory is contained in the Wilson coefficents entering the 
effective Hamiltonian which take different values in different theories with testable 
consequences in rare $B$ decays. 

There is an interesting class of rare radiative $B$-decays which proceed merely through the weak-annihilation mechanism. 
These processes have very small probabilities and have not been observed. 
So far, only upper limits on the branching ratios of these decays have been obtained: In 2004, the BaBar Collaboration 
provided the upper limit ${\cal B} (B^0\to J/\psi\gamma)<1.6\cdot 10^{-6}$ \cite{BaBar2004}. 
Very recently, the LHCb Collaboration reached the same sensitivity to the $B^0$-decay and set the limit on the $B_s^0$ decay: 
${\cal B} (B^0\to J/\psi\gamma)<1.7\cdot 10^{-6}$ and ${\cal B} (B_s^0\to J/\psi\gamma)<7.4\cdot 10^{-6}$ at 90\% CL 
\cite{LHCb2015}. 
Obviously, with the increasing statistics, the prospects to improve the limits on the branching ratios by one order of magnitude 
or eventually to observe these decays in the near future seem very favorable. 

The annihilation-type $B$-decays are promising from the perspective of obtaining theoretical predictions since the QCD dynamics 
of these decays is relatively simple \cite{gp,grinstein}. These decays have been addressed in the literature but---in spite 
of their relative 
simplicity---the available theoretical predictions turned out to be rather uncertain; for instance, the predictions for 
${\cal B} (B_s^0\to J/\psi\gamma)$ decay vary from $5.7\cdot 10^{-8}$ \cite{ch2004} to $5\cdot 10^{-6}$ \cite{ch2006}. 
The situation is clearly unsatisfactory and requires clarification. We did not find any of these results convincing and 
present in this paper a more detailed analysis of the $B\to V\gamma$ decays. 

\begin{center}
\begin{figure}
\mbox{\epsfig{file=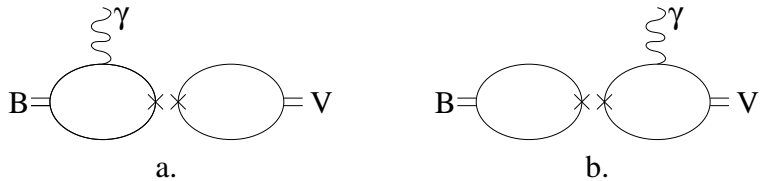,width=10cm}}
\caption{\label{fig:diag}
Diagrams describing the weak annihilation process for $B\to V\gamma$ in the factorization approximation: 
(a) The photon is emitted from the $B$-loop, 
(b) The photon is emitted from the vector-meson $V$-loop.}
\end{figure}
\end{center}

The annihilation type $B\to V\gamma$ decays proceed through the four-quark operators of the effective weak Hamiltonian. 
In the factorization approximation, the amplitude can be represented as the product of meson leptonic decay constants 
and matrix elements of the weak current between meson and photon; the latter contain the meson-photon transition form factors. 
The photon can be emitted from the loop containing the $B$-meson (Fig.~\ref{fig:diag}a), this contribution is described by 
the $B\gamma$ transition form factors. The photon can be also emitted from the vector-meson $V$-loop (Fig \ref{fig:diag}b); 
this contribution is described by the $V\gamma$ transition form factors. The latter were erroneously believed to give 
small contribution to the amplitude and have not been considered in the previous analyses.  
The main new ingredient of this paper is the analysis of the photon emission from the $V$-loop. 
First, we show that this contribution has no parametric suppression compared to the photon emission from the $B$-loop. 
Then, we calculate the $B\gamma$ and $V\gamma$ form factors within the relativistic dispersion approach based on 
the constituent quark picture 
\cite{m}. As shown in  \cite{melikhov}, the form factors from this approach satisfy all rigorous constrains which 
emerge in QCD in the 
limit of heavy-to-heavy and heavy-to-light transitions; as demonstrated in \cite{ms,bmns,mn2004}, the numerical results for the 
weak transition form factors from this approach exhibit an excellent agreement with the results from lattice QCD and 
QCD sum rules. 

Numerically, we report here that the $V$-loop contribution to the amplitude turns out to be comparable 
with the $B$-loop contribution and has a sizeable impact on the probability of the weak-annihilation $B\to V\gamma$ decay. 

The paper is organized as follows: In Section \ref{Sect2} the effective weak Hamiltonian and the structure of the amplitude are recalled. 
We discuss the general structure of the $B\to V\gamma$ amplitude and work out the constraints coming from gauge invariance.  
In Section \ref{Sect3} we consider the photon emission from the $B$-loop and present the $B\gamma$ transition 
form factors within the 
relativistic dispersion approach based on constituent quark picture. 
Section \ref{Sect4} contains the analysis of the $V\gamma$ transition form factors. 
Finally, in Section \ref{Sect5} the numerical estimates are given. 
The concluding Section \ref{Sect6} summarises our results and present a critical discussion of other results 
existing in the literature. 

\section{\label{Sect2}The effective Hamiltonian, the amplitude, and the decay rate}
We consider the weak-annihilation radiative $B\to V\gamma$ transition, where 
$V$ is the vector meson containing at least one charm quark, i.e. having the quark content $\bar q c$ ($q=u,d,s,c$). 
The corresponding amplitude is given by the matrix element of the effective Hamiltonian \cite{heff}
\begin{eqnarray}
A(B\to V\gamma)=\langle \gamma(q_1)V(q_2)|H_{\rm eff}|B(p) \rangle,  
\end{eqnarray}
where $p$ is the $B$ momentum, $q_2$ is the vector-meson momentum, and $q_1$ is the photon
momentum, $p=q_1+q_2$, $q_1^2=0$, $q_2^2=M_V^2$, $p^2=M_B^2$. 
The effective weak Hamiltonian relevant for the transition of interest has the form (we provide in this 
Section formulas for the effective Hamiltonian with the flavor structure 
$\bar d c\, \bar u b$, but all other decays of interest may be easily described by an obvious replacement of the 
quark flavors and the corresponding CKM factors $\xi_{\rm CKM}$):
\begin{eqnarray}
\label{Heff}
H_{\rm eff} &=& 
-\frac{G_F}{\sqrt{2}}{\xi_{\rm CKM}}
\left (C_1(\mu){\cal O}_1+C_2(\mu){\cal O}_2\right), 
\end{eqnarray}
$G_F$ is the Fermi constant, $\xi_{\rm CKM}=V^*_{cd}V_{ub}$, $C_{1,2}(\mu)$ are the scale-dependent Wilson coefficients \cite{heff}, 
and we only show the relevant four-quark operators 
\begin{eqnarray}
{\cal O}_1 &=& \bar d_{\alpha}\gamma_{\nu}(1-\gamma_5)c_{\alpha}\;
\bar u_{\beta}\gamma_{\nu}(1-\gamma_5) b_{\beta},\nonumber
\\
{\cal O}_2 &=& \bar d_{\alpha}\gamma_{\nu}(1-\gamma_5) c_{\beta}\; 
\bar u_{\beta}\gamma_{\nu}(1-\gamma_5) b_{\alpha}.
\end{eqnarray}
We use notations $e=\sqrt{4\pi\alpha_{\rm em}}$, 
$\gamma^5=i\gamma^0\gamma^1\gamma^2\gamma^3$,  
$\sigma_{\mu\nu}=i\left [\gamma_{\mu},\gamma_{\nu}\right ]/2$,
$\epsilon^{0123}=-1$ and
${\rm Sp}\left (\gamma^5\gamma^{\mu}\gamma^{\nu}\gamma^{\alpha}\gamma^{\beta}\right )
=4i\epsilon^{\mu\nu\alpha\beta}$.
 
The amplitude can be written as  
\begin{eqnarray}
\label{WA}
A(B\to V\gamma)&=&-\frac{G_F}{\sqrt{2}}\xi_{\rm CKM} a_{\rm eff}(\mu)
\langle V(q_2)\gamma(q_1)|\bar d\gamma_\nu(1-\gamma_5)u\cdot 
\bar u\gamma_\nu(1-\gamma_5)b|B(p)\rangle,  
\end{eqnarray}   
where $a_{\rm eff}(\mu)$ is an effective scale-dependent Wilson coefficient appropriate for the decay under consideration.

It is convenient to isolate the parity-conserving contribution which emerges from the product of the two equal-parity currents, and the parity-violating contribution 
which emerges from the product of the two opposite-parity currents. The amplitude may then be parameterized as follows  
\begin{eqnarray}
\label{F_PC}
A(B\to V\gamma)=\frac{eG_F}{\sqrt{2}}
\left[
\epsilon_{q_1\epsilon^\ast_1 q_2 \epsilon_2^\ast}F_{\rm PC}
+i \epsilon_2^{\ast\nu}\epsilon_1^{\ast\mu} \left(g_{\nu\mu}\,pq_1-p_\mu q_{1\nu}\right)F_{\rm PV}
\right], 
\end{eqnarray}
where $F_{\rm PC}$ and $F_{\rm PV}$ are the parity-conserving and 
parity-violating invariant amplitudes, respectively. Hereafter $\epsilon_2$($\epsilon_1$) is the 
vector-meson (photon) polarization vector. We use the short-hand notation 
$\epsilon_{abcd}=\epsilon_{\alpha\beta\mu\nu}a^{\alpha}b^{\beta}c^{\mu}d^{\nu}$ 
for any 4-vectors $a,b,c,d$. 

For the decay rate one finds 
\begin{eqnarray}
\label{rate}
\Gamma(B\to V\gamma)=\frac{G^2_F\,\alpha_{em}}{16}M_B^3
\left(1-{M^2_V}/{M_B^2}\right)^3
     \left( |F_{\rm PC}|^2+|F_{\rm PV}|^2 \right). 
\end{eqnarray}
Neglecting the nonfactorizable soft-gluon exchanges, i.e. assuming vacuum saturation, 
the complicated matrix element in Eq. (\ref{WA}) is reduced to simpler quantities - 
the meson-photon matrix elements of the bilinear quark currents and the 
meson decay constants. The latter are defined as usual  
\begin{eqnarray}
\langle V(q_2)|\bar d\gamma_\nu u|0\rangle &=& \epsilon_{2\nu}^\ast M_V f_V, \qquad  f_V>0,
\nonumber \\
\langle 0|\bar u\gamma_\nu \gamma_5 b|B(p)\rangle &=& ip_\nu f_B,\qquad  f_B>0.
\end{eqnarray}

\subsection{The parity-violating amplitude} 
The parity-violating contribution to the weak annihilation amplitude has the form  
\begin{eqnarray}
\label{apva}
A_{\rm PV}(B\to V\gamma)&=&\frac{G_F}{\sqrt{2}}\xi_{\rm CKM} a_{\rm eff}(\mu)  
\left\{
\langle V\gamma|\bar d \gamma_\nu u|0 \rangle 
\langle 0|\bar u \gamma_\nu\gamma_5 b|B \rangle  
+
\langle V|\bar d \gamma_\nu u|0 \rangle 
\langle \gamma|\bar u \gamma_\nu\gamma_5 b|B \rangle \right\}. 
\end{eqnarray}
It is convenient to denote 
\begin{eqnarray}
A_{\rm PV}^{(1)}=\langle V(q_2)|\bar d \gamma_\nu u|0 \rangle 
\langle \gamma(q_1)|\bar u \gamma_\nu\gamma_5 b|B(p) \rangle 
\end{eqnarray}
and 
\begin{eqnarray}
A_{\rm PV}^{(2)}=
\langle V(q_2)\gamma(q_1)|\bar d \gamma_\nu u|0 \rangle 
\langle 0|\bar u \gamma_\nu\gamma_5 b|B(p) \rangle.  
\end{eqnarray}

\noindent 1. Let us start with $A_{\rm PV}^{(1)}$. 
One can write  
\begin{eqnarray}
\langle \gamma(q_1)|\bar u \gamma_\nu\gamma_5 b|B(p) \rangle=
e\,\epsilon_1^{\ast\mu} T^B_{\mu\nu} 
\end{eqnarray}
where 
\begin{eqnarray}
T^B_{\mu\nu}(p,q_1)&=&i\int dx e^{iq_1x}\langle 0|T(J^{\rm e.m.}_\mu(x),\bar u\gamma_\nu \gamma_5 b)|B(p)\rangle,
\end{eqnarray}
and 
\begin{eqnarray}
J^{\rm e.m.}_\mu(x)=\frac23\left(\bar u\gamma_\mu u+\bar c\gamma_\mu c+\bar t\gamma_\mu t\right)
-\frac13 \left(\bar d\gamma_\mu d +\bar s\gamma_\mu s + \bar b\gamma_\mu b \right) 
\end{eqnarray}
is the electromagnetic quark current. 

The amplitude $T^B_{\mu\nu}$ in general contains 5 independent Lorentz structures 
and can be parameterized in various ways \cite{gp,mk,mn2004}. There is however the unique parameterization of the amplitude, 
which provides a distinct separation of the amplitude: form factors in the gauge-invariant transverse 
part of the amplitude, and contact terms in its longitudinal part \cite{m}: 
\begin{eqnarray}
\label{qq}
T^B_{\mu\nu}=T^{\perp}_{\mu\nu}+\frac{iq_{1\mu}p_\nu}{q^2_1} R_1+\frac{iq_{1\mu}q_{1\nu}}{q^2_1} R_2, 
\end{eqnarray}
with 
\begin{eqnarray}
\label{tperp}
T^{\perp}_{\mu\nu}&=&i\left(g_{\mu\nu}-\frac{q_{1\mu}q_{1\nu}}{q_1^2}\right)pq_1\,F_{A1}(q_1^2)
+i\left(p_{\mu}-\frac{pq_1}{q_1^2}\,q_{1\mu}\right)q_{1\nu}F_{A2}(q_1^2)
+i\left(p_{\mu}-\frac{pq_1}{q_1^2}\,q_{1\mu}\right)p_{\nu}F_{A3}(q_1^2).  
\end{eqnarray}
The invariant amplitudes $R_1$ and $R_2$ in the longitudinal structure can be determined using 
the conservation of the electromagnetic current $\partial_\mu J^{\rm e.m.}_\mu=0$ \cite{bmns}, which leads to  
\begin{eqnarray}
\label{eq3}
q_{1\mu} T^B_{\mu\nu}(p,q_1)&=&-\langle 0|[\hat Q, \bar u\gamma_\nu \gamma_5 b]|B(p)\rangle
=iQ_Bf_B p_\nu.  
\end{eqnarray}
and thus to  
\begin{eqnarray}
R_1&=&Q_Bf_B, \qquad
R_2=0. 
\end{eqnarray}
The parameterization (\ref{qq}) of the amplitude is prompted by the structure of 
the Feynman diagram: let us rewrite the usual electromagnetic coupling of the quark as follows ($q_1=k-k'$):
\begin{eqnarray}
(m+\hat k')\gamma_\mu(m+\hat k)
=(m+\hat k')\left\{\gamma_\mu-\hat q_1\,\frac{q_{1\mu}}{q_1^2} \right\}(m+\hat k)+
\frac{q_{1\mu}}{q_1^2}\,\left[(k^2-m^2) (m+\hat k')-(k'^2-m^2) (m+\hat k)\right].
\end{eqnarray} 
The first term is explicitly transverse with respect to $q_{1\mu}$ and leads to $T^\perp_{\mu\nu}$. 
The second term, containing the factors $(k^2-m^2)$ and $(k'^2-m^2)$, leads to the contact term 
$ip_\nu\frac{q_{1\mu}}{q_1^2}f_B$.  The Lorentz structures in (\ref{tperp}) have singularities at $q_1^2=0$, but 
the full amplitude $T_{\mu\nu}^B$ should be regular at $q_1^2=0$. So the singularities must cancel each 
other yielding the constraints on the form factors at $q_1^2=0$: 
\begin{eqnarray}
\label{relationsffs}
F_{A1}(0)=-F_{A2}(0), \qquad 
F_{A3}(0)=\frac{f_BQ_B}{pq_1}. 
\end{eqnarray}
Hereafter, when evaluating the invariant amplitudes at $q_1^2=0$, one should make use of relation $pq_1=\frac{1}{2}(M_B^2-M_V^2)$. 
By virtue of (\ref{relationsffs}), for the amplitude $A^{\rm PV}_1$ at $q_1^2=0$ we find  
\begin{eqnarray}
\label{A1PV}
A_{\rm PV}^{(1)}&=&
ie\,f_V M_V \epsilon_1^{\ast\mu}\epsilon_2^{\ast\nu}\left\{
g_{\mu\nu}pq_1 F_{A1}(0)
+p_{\mu}q_{1\nu}F_{A2}(0)
+p_\mu p_{\nu}F_{A3}(0)\right\}
\nonumber\\
&=&
i e\,f_V M_V \epsilon_1^{\ast\mu}\epsilon_2^{\ast\nu}\left\{
(g_{\mu\nu}pq_1-p_{\mu}q_{1\nu})\frac{F_A}{M_B}+p_\mu q_{1\nu}\frac{f_BQ_B}{pq_1}\right\}, 
\end{eqnarray}
with $F_A=M_BF_{A1}(0)$. Notice that the contact term does not contribute 
to the amplitude directly, but nevertheless determines the value of the form factor $F_{3A}(0)$. 

\noindent 2. Let us now turn to $A^{(2)}_{\rm PV}$. Using the equation of motion for the quark fields 
\begin{eqnarray}
\label{eom}
i\gamma_\nu\partial^\nu q(x)&=& m q(x) - Q_q A_\nu \gamma^\nu q(x), 
\nonumber\\
i\partial^\nu \bar q(x)\gamma_\nu&=& -m \bar q(x) + Q_q A_\nu \bar q(x)\gamma^\nu , 
\end{eqnarray}
one obtains 
\begin{eqnarray}
i\partial_\nu(\bar d \gamma_\nu c)=j+(Q_d-Q_c)\bar d \gamma_\nu c A_\nu, 
\end{eqnarray}
where 
\begin{eqnarray}
j(x)=(m_c-m_d)\bar d(x) c(x)
\end{eqnarray}
is the scale-independent scalar current. Then for the amplitude $A_{\rm PV}^{(2)}$ we find 
\begin{eqnarray}
A^{(2)}_{\rm PV}&=&i p_\nu f_B \langle V\gamma|\bar d \gamma_\nu c|0 \rangle
=
- i f_B \langle V\gamma|j|0 \rangle-i f_B (Q_d-Q_c)\epsilon_1^{\ast\nu}\langle V|\bar d \gamma_\nu c)|0 \rangle
\nonumber\\
&=&
- i f_B \langle V\gamma|j|0 \rangle-if_B \epsilon_1^{\ast\mu}\epsilon_2^{\ast\nu}g_{\mu\nu}Q_Vf_V M_V.  
\end{eqnarray}  
We have taken into account here the charge-conservation relation 
\begin{eqnarray}
Q_b-Q_u=Q_B=Q_V=Q_d-Q_c. 
\end{eqnarray}  
The amplitude $\langle V\gamma|j|0 \rangle$ may be written as
\begin{eqnarray}
\langle V(q_2)\gamma(q_1)|j|0 \rangle
=
i e \epsilon_1^{\ast\mu}(q_1)\langle V(q_2)|\int dx e^{i q_1 x}T(J^{\rm e.m.}_\mu(x) j(0))|0 \rangle
\equiv i e \epsilon_1^{\ast\mu}(q_1)T_\mu^V, 
\end{eqnarray}  
and for $T_\mu^V$ one can write the decomposition  
\begin{eqnarray}
T_\mu^V=i \epsilon_2^{\ast\nu}(q_2)
\left\{\left(g_{\mu\nu}-\frac{q_{1\mu}q_{1\nu}}{q_1^2}\right)pq_1\,H_{S1}(q_1^2)
+\left(p_{\mu}-\frac{pq_1}{q_1^2}\,q_{1\mu}\right)q_{1\nu}H_{S2}(q_1^2)\right\}.
\end{eqnarray} 
Making use of the electromagnetic current conservation, one finds that the contact terms in $T_\mu^V$ are absent
due to the relation $\langle V|j|0\rangle=0$. Again, the singularities at $q_1^2=0$ of the transverse Lorentz projectors should cancel in the amplitude which is free from the singularity at $q_1^2=0$, leading to
\begin{eqnarray}
H_{S1}(0)=-H_{S2}(0).  
\end{eqnarray} 
Then, for the radiative decay $q_1^2=0$, 
one obtains 
\begin{eqnarray}
\langle V(q_2)\gamma(q_1)|j|0 \rangle=i e f_B\epsilon_1^{\ast\mu}\epsilon_2^{\ast\mu}
\left(g_{\mu\nu}pq_1-p_\mu q_{1\nu}\right)H_S, 
\end{eqnarray}  
with $H_S=H_{S1}(0)$. 
Finally, using charge conservation $Q_V=Q_B$, we arrive at 
\begin{eqnarray}
\label{A2PV}
A^{(2)}_{\rm PV}=i e f_B\epsilon_1^{\ast\mu}\epsilon_2^{\ast\nu}
\left(g_{\mu\nu}pq_1-p_\mu q_{1\nu}\right)H_S-if_B\epsilon_1^{\ast\mu}\epsilon_2^{\ast\mu}g_{\mu\nu}Q_Vf_V M_V.
\end{eqnarray}  

\noindent 3. We are ready now to obtain the parity-violating contribution to the amplitude in the factorization approximation. First, let us mention that making use of the charge-conservation $Q_B=Q_V$, the sum of the separately gauge-noninvariant terms in (\ref{A1PV}) and (\ref{A2PV}) yields a gauge-invariant combination 
\begin{eqnarray}
-ie \epsilon_1^{\ast\mu}\epsilon_2^{\ast\mu} \left(g_{\mu\nu}-\frac{p_\mu q_{1\nu}}{pq_1}\right)f_B Q_B {f_V M_V}. 
\end{eqnarray} 
For the sum $A_{\rm PV}^{(1)}+A_{\rm PV}^{(2)}$, we then find an explicitly gauge-invariant expression 
\begin{eqnarray}
A_{\rm PV}^{(1)}+A_{\rm PV}^{(2)}&=&ie\,\epsilon_1^{\ast\mu}\epsilon_2^{\ast\nu}
\left(g_{\mu\nu}pq_1-p_{\mu}q_{1\nu}\right)\left[\frac{F_{A}}{M_B}f_V M_V+f_B H_S
-\frac{Q_Bf_Bf_VM_V}{pq_1}\right],
\end{eqnarray}  
such that the parity-violating amplitude of (\ref{F_PC})
\begin{eqnarray}
F_{\rm PV}=\xi_{\rm CKM}a_{\rm eff}(\mu)\left[\frac{F_{A}}{M_B}f_V M_V+f_B H_S
-\frac{2Q_Bf_Bf_VM_V}{M_B^2-M_V^2}\right].
\end{eqnarray}  


\subsection{The parity-conserving amplitude} 
This amplitude reads  
\begin{eqnarray}
\label{apc}
A_{\rm PC}(B\to V\gamma)&=&-\frac{G_F}{\sqrt{2}}\xi_{\rm CKM} a_{\rm eff}(\mu) 
\left\{\langle V|\bar d\gamma_\nu u|0 \rangle \langle \gamma|\bar u \gamma_\nu b|B \rangle 
+
\langle \gamma V|\bar d\gamma_\nu \gamma_5 u |0 \rangle 
\langle 0| \bar u\gamma_\nu \gamma_5 b|B \rangle \right\}. 
\end{eqnarray}
\noindent 1. The first contribution to the amplitude, corresponding to the photon emission from the $B$-meson loop, reads  
\begin{eqnarray}
A^{(1)}_{\rm PC}=\langle V|\bar d\gamma_\nu c|0 \rangle \langle \gamma|\bar u \gamma_\nu b|B \rangle
=-e M_V f_V \epsilon_{q_1\epsilon^{\ast}_1 q_2 \epsilon^{\ast}_2}\;\frac{F_V}{M_B},  
\end{eqnarray}
where $F_V$ is the form factor describing the $B\to\gamma$ transition induced by the vector weak current     
\begin{eqnarray}
\langle \gamma(q_1)|\bar u\gamma_\nu b|B(p)\rangle = 
-e\,\epsilon_{q_1\epsilon_1^{\ast} q_2\nu}\frac{F_V}{M_B}.
\end{eqnarray}
\noindent 2. The second term in (\ref{apc}), describing the photon emission from the vector-meson loop, 
may be reduced to the divergence of the axial-vector current. Making use of the equations of motion (\ref{eom}) one finds
\begin{eqnarray}
i\partial_\nu(\bar d\gamma_\nu\gamma_5 c)=-j^5+(Q_d-Q_c)\bar d\gamma_\nu\gamma_5 c A^\nu
\end{eqnarray}
with the scale-independent pseudoscalar current 
\begin{eqnarray}
\label{j5}
j^5=(m_d+m_c)\bar d\gamma_5 c. 
\end{eqnarray}
Taking into account that $\langle V|\bar d\gamma_\nu\gamma_5 c|0 \rangle=0$, we find   
\begin{eqnarray}
A^{(2)}_{\rm PC}=\langle 0|\bar u\gamma_\nu \gamma_5 b|B\rangle 
\langle \gamma V|\bar d\gamma_\nu\gamma_5 c|0 \rangle&=&
f_B \langle \gamma V|\partial_\nu (\bar d\gamma_\nu\gamma_5 c) |0 \rangle=
-e f_B \epsilon_{q_1\epsilon_1^\ast q_2\epsilon_2^\ast}H_P,
\end{eqnarray}
where the form factor $H_P$ is defined as  
\begin{eqnarray}
\label{defHp}
\langle \gamma(q_1) V(q_2)|j_5|0 \rangle=i e\epsilon_{q_1\epsilon_1^\ast q_2\epsilon_2^\ast}H_P. 
\end{eqnarray}
\noindent 3. Finally, the parity-conserving invariant amplitude of (\ref{F_PC}) takes the form  
\begin{eqnarray}
\label{fpc}
F_{\rm PC}=\xi_{\rm CKM} a_{\rm eff}(\mu)\left[\frac{F_V}{M_B}f_VM_V +f_B H_P\right]. 
\end{eqnarray}

Summing up this Section, within the factorization approximation the weak annihilation amplitude may be expressed in terms of four form factors: 
$F_A$, $F_V$, $H_P$ and $H_S$. 
It should be emphasized that each of the form factors $F_A$, $F_V$, $H_P$ and $H_S$ actually depends on two variables: 
The $B$-meson transition form factors $F_A$, $F_V$ depend on $q_1^2$ and $q_2^2$, and  
$F_{A,V}(q_1^2,q_2^2)$ should be evaluated at $q_1^2=0$ and $q_2^2=M_V^2$. 
The vector-meson transition form factors $H_P$ and $H_S$ 
depend on $q_1^2$ and $p^2$, and $H_{S,P}(q_1^2,p^2)$ should be evaluated at $q_1^2=0$ and $p^2=M_B^2$.

\section{\label{Sect3}Photon emission from the $B$-meson loop and the form factors $F_A$ and $F_V$.}
In this section we calculate the form factors $F_{A,V}$ within the relativistic dispersion approach to the transition form factors based 
on constituent quark picture. This approach has been formulated in detail in 
\cite{melikhov} and applied to the weak decays of heavy mesons in \cite{ms}. 

The pseudoscalar meson in the initial state is described in the dispersion approach by the following vertex \cite{m}: 
$\bar q_1(k_1)\; i\gamma_5 q(-k_2)\;G(s)/{\sqrt{N_c}}$, 
with $G(s)=\phi_P(s)(s-M_P^2)$, $s=(k_1+k_2)^2$, $k_1^2=m_1^2$ and $k_2^2=m_2^2$. 
The pseudoscalar-meson wave function $\phi_P$ is normalized according to the relation \cite{m}
\begin{eqnarray}
\label{norma}
\frac{1}{8\pi^2}\int\limits_{(m_1+m_2)^2}^\infty ds \phi_P^2(s)
\left({s-(m_1-m_2)^2}\right)\frac{\lambda^{1/2}(s,m_1^2,m_2^2)}{s}=1.  
\end{eqnarray}
The decay constant is represented through $\phi_P(s)$ by the spectral integral
\begin{eqnarray}
\label{fP}
f_P=\sqrt{N_c}\int\limits_{(m_1+m_2)^2}^\infty ds \phi_P(s)
(m_1+m_2)\frac{\lambda^{1/2}(s,m_1^2,m_2^2)}{8\pi^2s}\frac{s-(m_1-m_2)^2}{s}. 
\end{eqnarray}
Here $\lambda(a,b,c)=(a-b-c)^2-4bc$ is the triangle function.

Recall that the form factors $F_{A,V}$ describe the transition of the $B$-meson to the photon with the momentum 
$q_1$, $q_1^2=0$, induced by the axial-vector (vector) current with the momentum $q_2$, $q_2^2=M_V^2$. 
We derive the double spectral representations for the form factor in $p^2$ and $q_2^2$; this allows us to avoid the appearance 
of the unphysical polynomial terms in the amplitudes which otherwise should be killed by appropriate subtractions.

\subsection{The form factor $F_A$}
\begin{center}
\begin{figure}[h!]
\mbox{\epsfig{file=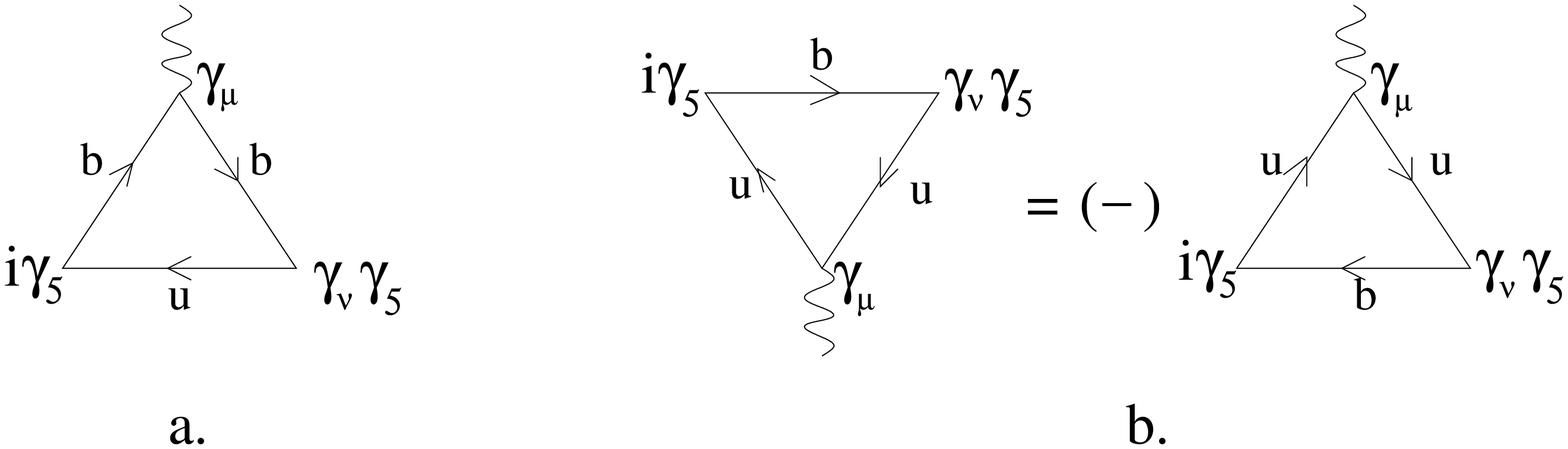,width=13cm}}
\caption{\label{fig:Fa}Diagrams for the form factor $F_A$: a) $F_A^{(b)}$, b) $F_A^{(u)}$.} 
\mbox{\epsfig{file=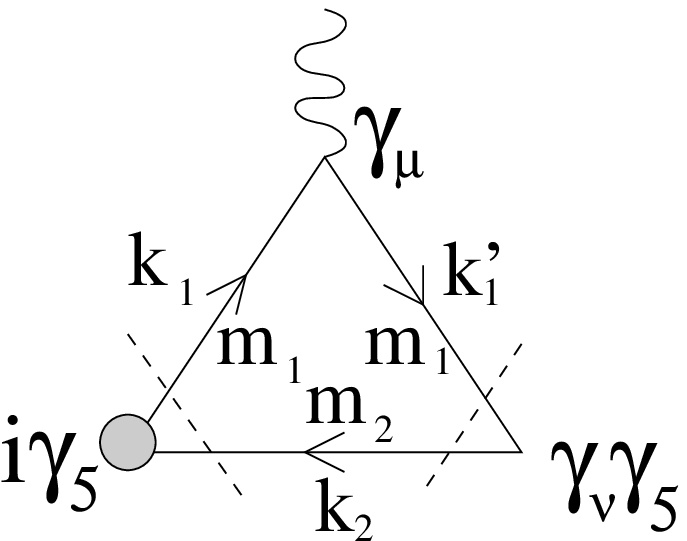,width=4cm}}
\caption{\label{fig:Fat} 
The triangle diagram for $F_A^{(1)}(m_1,m_2)$. The cuts correspond to calculating the double spectral 
density in $p^2$ and $q_2^2$.}
\end{figure}
\end{center}

The form factor $F_A$ is given by the diagrams of Fig \ref{fig:Fa}. 
Fig \ref{fig:Fa}a shows $F_A^{(b)}$, the contribution to the form factor of the process when the $b$ quark interacts with the 
photon; Fig \ref{fig:Fa}b describes the contribution of the process when the quark $u$ interacts while 
$b$ remains a spectator. 

It is convenient to change the direction of the quark line in the loop diagram of Fig \ref{fig:Fa}b. This is done by performing the charge conjugation of the matrix element and leads to a sign change for the $\gamma_\nu\gamma_5$ vertex. Now both diagrams in Fig \ref{fig:Fa} 
a,b are reduced to the diagram of Fig \ref{fig:Fat} which defines the form factor $F_A^{(1)}(m_1,m_2)$: 
Setting $m_1=m_b$, $m_2=m_u$ gives $F_A^{(b)}$, while 
setting $m_1=m_u$, $m_2=m_b$ gives $F_A^{(u)}$ such that      
\begin{eqnarray}
F_A=Q_b F_A^{(b)}-Q_u F_A^{(u)}.
\end{eqnarray} 
For the diagram of Fig \ref{fig:Fat} (quark 1 emits the photon, quark 2 is the spectator, all quark lines are on their mass shell) the trace reads  
\begin{eqnarray}
-{\rm Sp}\left(i\gamma_5(m_2-\hat k_2)\gamma_\nu\gamma_5(m_1+\hat k'_1)\gamma_\mu(m_1+\hat k_1)\right)&=&
4i(k_1+k_1')_\mu(m_1 k_2+m_2 k_1)_\nu\nonumber\\
&&+ 4i(g_{\mu\nu}q_{1\alpha}-g_{\mu\alpha}q_{1\nu})(m_1 k_2+m_2 k_1)_\alpha. 
\end{eqnarray}
The double spectral density of the form factor $F^{(1)}_A(m_1,m_2)$ in the variables $p^2$, $p=k_1+k_2$, and $q_2^2$, 
$q_2=k'_1+k_2$, is obtained as the coefficient of the structure $g_{\mu\nu}$ after the integration of the trace over the quark phase space. 
At $q_1^2=0$, the double spectral representation for the elastic form factor is reduced to a single spectral representation, which is given below. 

The easiest was to derive this spectral representation is to use the light-cone variables \cite{amn}. 
Performing the necessary calculations, we arrive at the following representation
\begin{eqnarray}
\label{fa-lc}
\frac{1}{M_B}F_{A}^{(1)}(m_1,m_2)&=&\frac{\sqrt{N_c}}{4\pi^2}\int \frac{dx_1 dx_2 dk_\perp^2}{x_1^2 x_2}\delta(1-x_1-x_2)
\frac{\phi_B(s)}{s-M_V^2}\left(m_1x_2+m_2 x_1+(m_1-m_2)\frac{2k_\perp^2}{M_B^2-M_V^2}\right).  
\end{eqnarray}
Here $x_i$ is the fraction of the $B$-meson light-cone momentum carried by the quark $i$, and 
\begin{eqnarray}
\label{s}
s=\frac{m_1^2}{x_1}+\frac{m_2^2}{x_2}+\frac{k_\perp^2}{x_1x_2}. 
\end{eqnarray}
This expression may be cast in the form of a single dispersion integral 
\begin{eqnarray}
\label{fadisp}
\frac{1}{M_B}F_{A}^{(1)}(m_1,m_2)&=&\frac{\sqrt{N_c}}{4\pi^2}\int\limits_{(m_1+m_2)^2}^\infty
\frac{ds\;\phi_B(s)}{(s-M_V^2)}
\left(\rho_+(s,m_1,m_2)+2\frac{m_1-m_2}{M_B^2-M_V^2}\rho_{k_\perp^2}(s,m_1,m_2)\right), 
\end{eqnarray}
where 
\begin{eqnarray}
\label{rhoplus}
\rho_+(s,m_1,m_2)&=&(m_2-m_1)\frac{\lambda^{1/2}(s,m_1^2,m_2^2)}{s}+m_1\log\left(\frac{s+m_1^2-m_2^2+\lambda^{1/2}(s,m_1^2,m_2^2)}
{s+m_1^2-m_2^2-\lambda^{1/2}(s,m_1^2,m_2^2)}\right),
\\
\label{rhokperp2}
\rho_{k_\perp^2}(s,m_1,m_2)&=&\frac{s+m_1^2-m_2^2}{2s}\lambda^{1/2}(s,m_1^2,m_2^2)-
m_1^2\log\left(\frac{s+m_1^2-m_2^2+\lambda^{1/2}(s,m_1^2,m_2^2)}
{s+m_1^2-m_2^2-\lambda^{1/2}(s,m_1^2,m_2^2)}\right).
\end{eqnarray}
Making use of the light-cone representation (\ref{fa-lc}), the light-cone representation of the pseudoscalar-meson decay constant (\ref{fP}) 
\begin{eqnarray}
f_P=\frac{\sqrt{N_c}}{4\pi^2}\int \frac{dx_1 dx_2dk_\perp^2}{x_1 x_2}\delta(1-x_1-x_2)
\delta\left(s-\frac{m_1^2}{x_1}-\frac{m_2^2}{x_2}-\frac{k_\perp^2}{x_1x_2}\right)\phi_P(s)\left(m_1 x_2 +m_2 x_1\right), 
\end{eqnarray}
and employing the fact that the wave function $\phi_P(s)$ is localized near the threshold in the region $\sqrt{s}-m_b-m_u\le \bar\Lambda$, it is easy to show that in the limit $m_b\to\infty$ the photon emission from the light quark dominates over the emission from the heavy quark \cite{korch}
\begin{eqnarray}
\label{Fa_HQ}
\frac{1}{M_B}F_{A}^{(u)}=\frac{f_B}{\bar\Lambda m_b}+...,\qquad  \frac{1}{M_B}F_{A}^{(b)}=\frac{f_B}{m_b^2}+...
\end{eqnarray}

\subsection{The form factor $F_V$} 
\begin{center}
\begin{figure}[ht]
\mbox{\epsfig{file=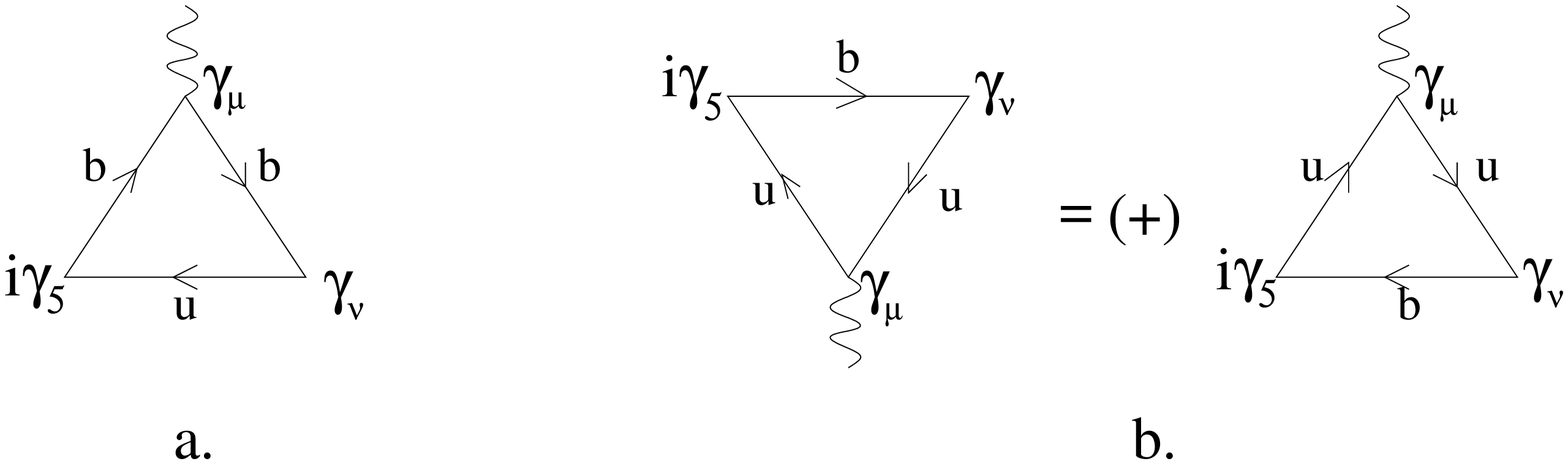,width=13cm}}
\caption{\label{fig:Fv}Diagrams for the form factor $F_V$: a) $F_V^{(b)}$, b) $F_V^{(u)}$.} 
\mbox{\epsfig{file=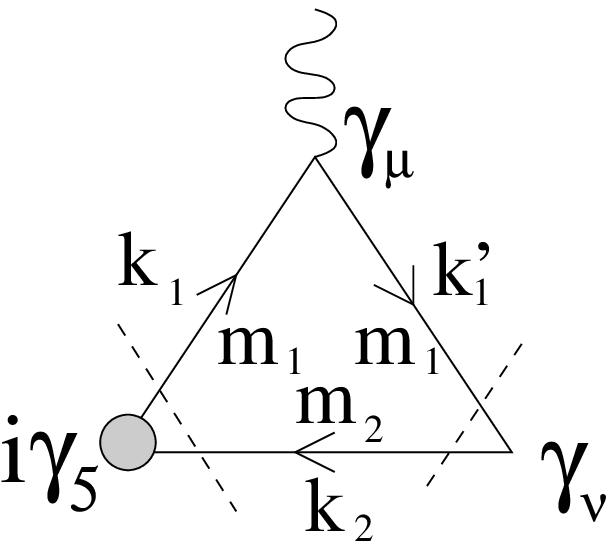,width=4cm}}
\caption{\label{fig:Fvt} The triangle diagram for $F_V^{(1)}(m_1,m_2)$. 
The cuts correspond to calculating the double spectral density in $p^2$ and $q_2^2$.}
\end{figure}
\end{center}

The consideration of the form factor $F_{V}$ is very similar to the form factor $F_{A}$. 
$F_V$ is determined by the two diagrams shown in Fig \ref{fig:Fv}:  
Fig \ref{fig:Fv}a gives $F_V^{(b)}$, the contribution of the process when the $b$ quark interacts with the photon; Fig \ref{fig:Fv}b 
describes the contribution of the process when the quark $u$ interacts. 

It is again convenient to change the direction of the quark line in the loop diagram of Fig \ref{fig:Fv}b by performing the charge 
conjugation of the matrix element. 
For the vector current $\gamma_\nu$ in the vertex the sign does not change (in contrast to the $\gamma_\nu\gamma_5$ case considered above). 
Then both diagrams in Fig \ref{fig:Fv} a, b are reduced to the diagram of Fig \ref{fig:Fvt} which gives the form factor $F_V^{(1)}(m_1,m_2)$: 
Setting $m_1=m_b$, $m_2=m_u$ gives $F_V^{(b)}$ while setting $m_1=m_u$, $m_2=m_b$ gives $F_V^{(u)}$ such that  
\begin{eqnarray}
F_V=Q_bF_V^{(b)}+Q_uF_V^{(u)}. 
\end{eqnarray}
The trace corresponding to the diagram of Fig 4 (1 - active quark, 2 - spectator) reads 
\begin{eqnarray}
-{\rm Sp}\left(
i\gamma_5(m_2-\hat k_2)\gamma_\nu (m_1+\hat k'_1)\gamma_\mu(m_1+\hat k_1)\right)=
4\epsilon_{\mu q_1 \nu \alpha}(m_1 k_2+m_2 k_1)_\alpha. 
\nonumber
\end{eqnarray}
The light-cone representation of the form factor corresponding to \ref{fig:Fvt} takes the form 
\begin{eqnarray}
\label{fv-lc}
\frac{1}{M_B}F_{V}^{(1)}(m_1,m_2)&=&-\frac{\sqrt{N_c}}{4\pi^2}
\int \frac{dx_1 dx_2 dk_\perp^2}{x_1^2 x_2}\delta(1-x_1-x_2)
\frac{\phi_B(s)}{s-M_V^2}\left(m_1x_2+m_2 x_1\right),   
\end{eqnarray}
which may be written as a single spectral integral 
\begin{eqnarray}
\frac{1}{M_B}F_V^{(1)}(m_1,m_2)=-\frac{\sqrt{N_c}}{4\pi^2}\int\limits_{(m_1+m_2)^2}^\infty
\frac{ds\phi_B(s)}{(s-M_V^2)}\rho_+(s,m_1,m_2). 
\end{eqnarray}
The function $\rho_+(s,m_1,m_2)$ is given in (\ref{rhoplus}). 
In the heavy-quark limit $m_b\to\infty$ one finds 
\begin{eqnarray}
\label{Fv_HQ}
\frac{1}{M_B}F_{V}^{(u)}=-\frac{f_B}{\bar\Lambda m_b}+..., \qquad \frac{1}{M_B}F_{V}^{(b)}=-\frac{f_B}{m_b^2}+...
\end{eqnarray}
The dominant contribution in the heavy quark limit again comes from the process 
when the light quark emits the photon. As seen from Eqs. (\ref{Fa_HQ}) and (\ref{Fv_HQ}), 
one finds $F_A=F_V$ in the heavy quark limit, in agreement with the large-energy effective theory \cite{leet}. 

\section{\label{Sect4}Photon emission from the vector meson loop. The form factors $H_S$ and $H_P$.}
We now calculate the form factors $H_{P,S}$ using the relativistic dispersion approach. 
The vector meson in the final state is described in this approach by the vertex $\bar q_2(-k_2)\Gamma_\beta q_1(k_1')$, 
$\Gamma_\beta=\left(-\gamma_\beta+\frac{(k_1'-k_2)_\beta}{\sqrt{s}+m_1+m_2}\right)\;G(s)/{\sqrt{N_c}}$, 
with $G(s)=\phi_V(s)(s-M_V^2)$, $s=(k'_1+k_2)^2$, ${k'}_1^2=m_1^2$ and $k_2^2=m_2^2$. 
The vector-meson wave function $\phi_V$ is normalized according to \cite{melikhov}
\begin{eqnarray}
\label{normaV}
\frac{1}{8\pi^2}\int\limits_{(m_1+m_2)^2}^\infty ds \phi_V^2(s)
\left({s-(m_1-m_2)^2}\right)\frac{\lambda^{1/2}(s,m_1^2,m_2^2)}{s}=1.  
\end{eqnarray}
Its decay constant is represented through $\phi_V(s)$ by the spectral integral
\begin{eqnarray}
\label{fV}
f_V=\sqrt{N_c}\int\limits_{(m_1+m_2)^2}^\infty ds \phi_V(s)
\frac{2\sqrt{s}+m_1+m_2}{3}\frac{\lambda^{1/2}(s,m_1^2,m_2^2)}{8\pi^2s}\frac{s-(m_1-m_2)^2}{s}. 
\end{eqnarray}
Now the form factors $H_{S,P}$ describe the transition of the current with momentum $p$, $p^2=M_B^2$, 
to the photon with momentum 
$q_1$, $q_1^2=0$, and the vector meson with the momentum $q_2$, $q_2^2=M_V^2$. 
Similar to the previous section, we derive the double spectral representations for the form factor in $p^2$ and $q_2^2$. 

\subsection{The form factor $H_S$}
\begin{center}
\begin{figure}[b!]
\mbox{\epsfig{file=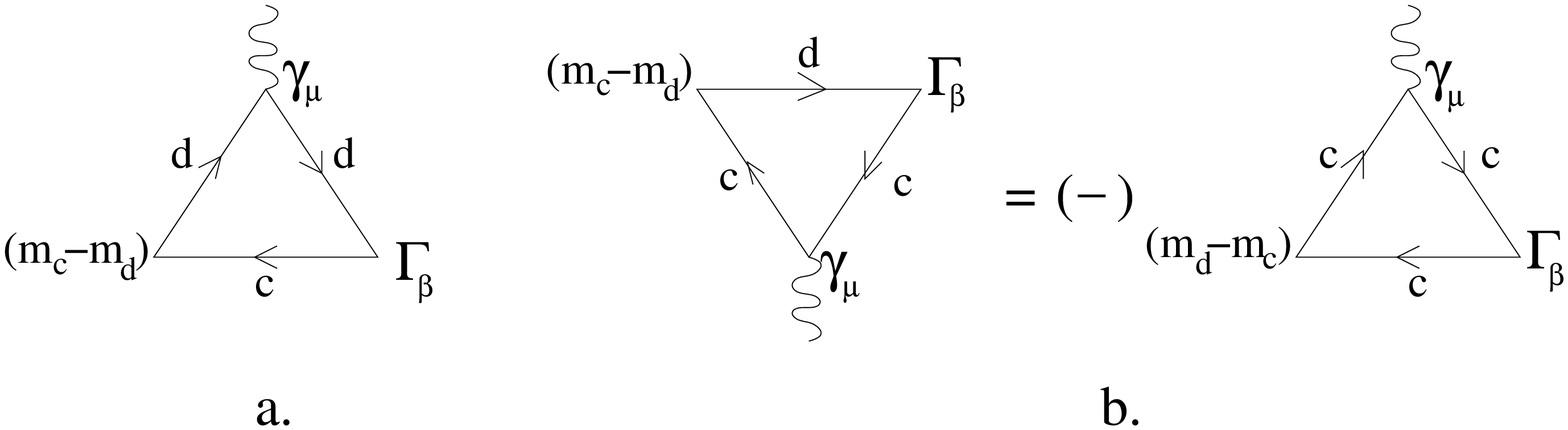,width=13cm}} 
\caption{\label{fig:Hs}Diagrams for the form factor $H_S$: a) $H_S^{(d)}$, b) $H_S^{(c)}$.} 
\mbox{\epsfig{file=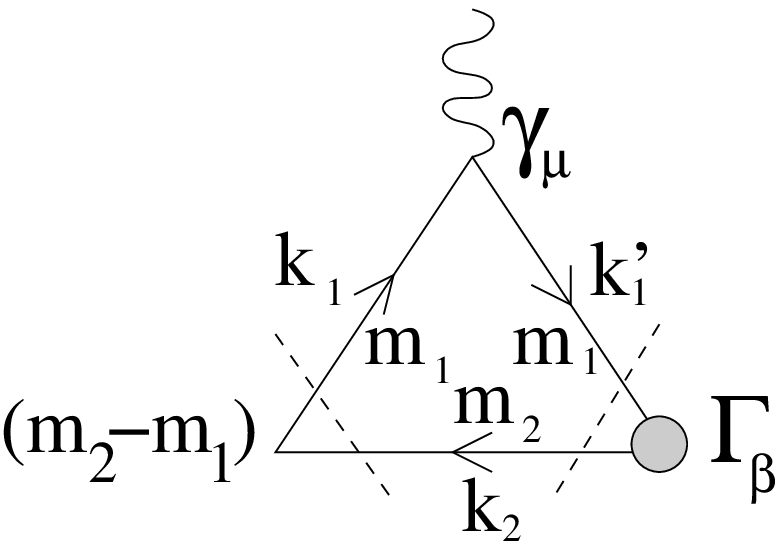,width=4cm}}
\caption{\label{fig:Hst} The triangle diagram for $H_S^{(1)}(m_1,m_2)$.}
\end{figure}
\end{center}

The form factor $H_S$ is given by the diagrams of Fig \ref{fig:Hs}. 
Fig \ref{fig:Fa}a shows $H_S^{(d)}$, the contribution to the form factor of the process when the $d$ quark interacts with the 
photon; Fig \ref{fig:Hs}b describes the contribution of the process when the quark $u$ interacts while 
$d$ remains spectator. 
Changing the direction of the quark line in the loop diagram of 
Fig \ref{fig:Hs}b, leads to a sign change for the scalar current $j=(m_c-m_d)\bar d c$ in the vertex, 
such that both diagrams in Fig \ref{fig:Hs}a,b are reduced to the diagram of Fig \ref{fig:Hst} which defines 
the form factor $H_S^{(1)}(m_1,m_2)$: 
Setting $m_1=m_d$, $m_2=m_c$ gives $H_S^{(d)}$, while 
setting $m_1=m_c$, $m_2=m_d$ gives $H_S^{(c)}$ such that      
\begin{eqnarray}
H_S=Q_d H_S^{(d)}-Q_c H_S^{(c)}.
\end{eqnarray} 
For the diagram of Fig \ref{fig:Hst} (quark 1 emits the photon, quark 2 is the spectator, 
all quark lines are on their mass shell) 
the trace for $q_1^2=0$ reads   
\begin{eqnarray}
&&-{\rm Sp}\left((m_1+\hat k'_1)\gamma_\mu(m_1+\hat k_1)(m_2-\hat k_2)\Gamma_\beta\right)=\\
\nonumber
&&-4(k_1+k_1')_\mu(m_1 k_2-m_2 k_1)_\beta-4(g_{\mu\beta}q_\alpha-g_{\mu\alpha}q_\beta)(m_1 k_2-m_2 k_1)_\alpha
+2\frac{(k'_1-k_2)_\beta}{\sqrt{s}+m_1+m_2}(k_1+k_1')_\mu(s-(m_1+m_2)^2). 
\end{eqnarray}
The double spectral density of $H_S^{(1)}(m_1,m_2)$ in $p^2$, $p=k_1+k_2$, and $q_2^2$, 
$q_2=k'_1+k_2$, is obtained as the coefficient of the structure $g_{\mu\nu}$ after the integration of the trace over 
the quark phase space. The light-cone representation for the form factor reads
\begin{eqnarray}
\label{hs-lc}
H_S^{(1)}(m_1,m_2)&=&\frac{\sqrt{N_c}}{4\pi^2}\int \frac{dx_1 dx_2 dk_\perp^2}{x_1^2 x_2}\delta(1-x_1-x_2)
\frac{\phi_V(s)}{s-p^2-i0}(m_2-m_1)\left(m_1x_2-m_2 x_1+\frac{2k_\perp^2\sqrt{s}}{p^2-M_V^2}\right).
\end{eqnarray}
with $s$ given in terms of $x_{1,2}$ and $k_\perp^2$ by (\ref{s}).
The corresponding single dispersion integral has the form 
\begin{eqnarray}
\label{hsdisp}
H_S^{(1)}(m_1,m_2)&=&\frac{\sqrt{N_c}}{4\pi^2}\int\limits_{(m_1+m_2)^2}^\infty
\frac{ds\;\phi_V(s)}{(s-p^2-i0)}(m_2-m_1)
\left(\rho_+(s,m_1,-m_2)+\frac{2\sqrt{s}}{p^2-M_V^2}\rho_{k_\perp^2}(s,m_1,m_2)\right), 
\end{eqnarray}
where $\rho_+(s,m_1,m_2)$ and $\rho_{k_\perp^2}(s,m_1,m_2)$ are determined earlier in (\ref{rhoplus}) and (\ref{rhokperp2}).

The light-cone representation (\ref{hs-lc}) allows us to obtain the behaviour of the form factor in the limit $m_Q\to\infty$ 
for the heavy-light vector meson $\bar Qq$ (and assuming $p^2\sim m_Q^2$): 
\begin{eqnarray}
\label{Hs_HQ}
H_S^{(q)}\propto f_V/\bar \Lambda,\qquad H_S^{(Q)}\propto f_V/m_Q,
\end{eqnarray}
but one expects a strong numerical suppression because of the partial cancellation of the leading-order contributions.

\subsection{The form factor $H_P$} 
\begin{center}
\begin{figure}[b]
\mbox{\epsfig{file=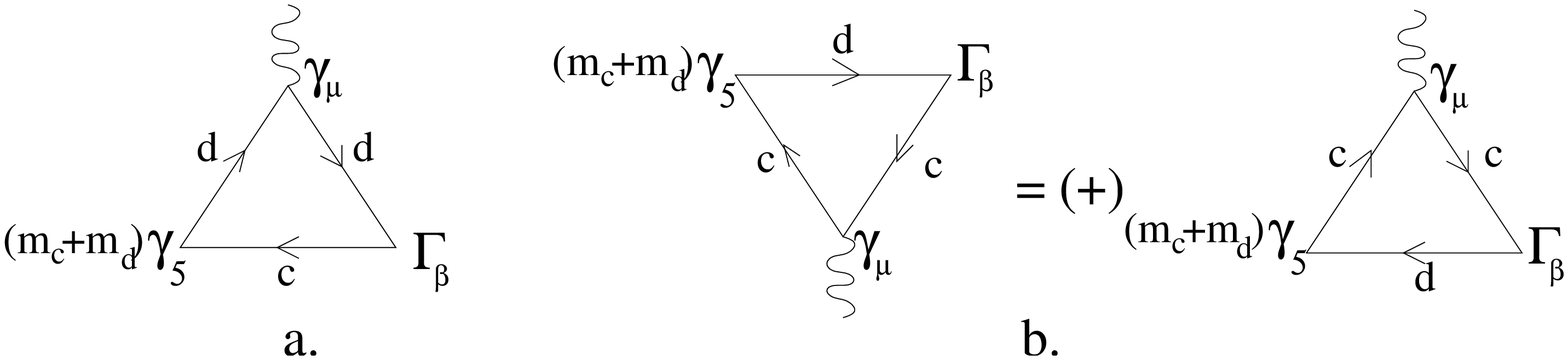,width=13cm}}
\caption{\label{fig:Hp}Diagrams for the form factor $H_P$: a) $H_P^{(d)}$, b) $H_P^{(u)}$.} 
\mbox{\epsfig{file=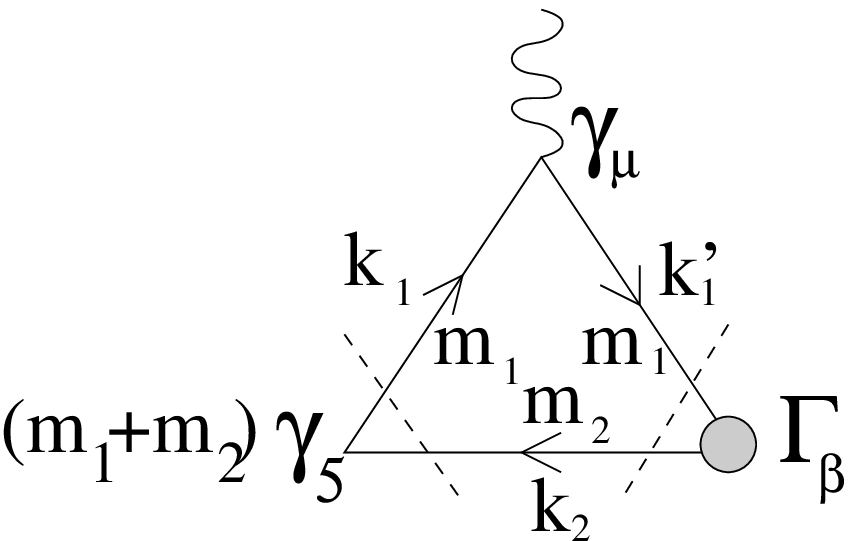,width=4cm}}
\caption{\label{fig:Hpt} The triangle diagram for $H_P^{(1)}(m_1,m_2)$. 
The cuts correspond to calculating the double spectral density in $p^2$ and $q_2^2$.}
\end{figure}
\end{center}

The form factor $H_{P}$ is determined by the two diagrams shown in Fig \ref{fig:Hp}:  
Fig \ref{fig:Hp}a gives $H_P^{(d)}$, the contribution of the process when the $d$-quark interacts with the 
photon; Fig \ref{fig:Hp}b describes the contribution of the process when the $c$-quark interacts. 
We again change the direction of the quark line in the loop diagram of 
Fig \ref{fig:Hp}b by performing the charge conjugation of the matrix element. 
For the pseudoscalar current $(m_c+m_d)\bar d\gamma_5 c$ in the vertex the sign does not change and both 
diagrams in Fig.~\ref{fig:Hp} a, b are reduced to the diagram of Fig \ref{fig:Hpt} which gives the 
form factor $H_P^{(1)}(m_1,m_2)$: Setting $m_1=m_d$, $m_2=m_c$ gives $H_P^{(d)}$ while  
setting $m_1=m_c$, $m_2=m_d$ gives $H_P^{(c)}$ such that  
\begin{eqnarray}
H_P=Q_dH_P^{(d)}+Q_cH_P^{(c)}. 
\end{eqnarray}
The trace corresponding to the diagram of Fig.~\ref{fig:Hpt} (1 - active quark, 2 - spectator) reads 
\begin{eqnarray}
-{\rm Sp}\left(
(m_1+\hat k'_1)\gamma_\mu(m_1+\hat k_1)\gamma_5(m_2-\hat k_2)\Gamma_\beta\right)=
4i\epsilon_{\mu q_1 \beta q_2}(m_1 k_2+m_2 k_1)_\alpha+
4i\epsilon_{\mu q_1 \alpha q_2}\frac{k_{2\alpha}(k_1'-k_2)_\beta}{\sqrt{s}+m_1+m_2}. 
\nonumber
\end{eqnarray}
The light-cone representation of the form factor corresponding to the diagram of Fig.~\ref{fig:Fvt} takes the form 
\begin{eqnarray}
\label{hp-lc}
H_P^{(1)}(m_1,m_2)&=&\frac{\sqrt{N_c}}{4\pi^2}
\int \frac{dx_1 dx_2 dk_\perp^2}{x_1^2 x_2}\delta(1-x_1-x_2)
\frac{\phi_V(s)}{s-p^2-i0}(m_1+m_2)\left(m_1x_2+m_2 x_1+\frac{k_\perp^2}{\sqrt{s}+m_1+m_2}\right),\nonumber\\   
\end{eqnarray}
which may be written as a single spectral integral 
\begin{eqnarray}
H_P^{(1)}(m_1,m_2)=\frac{\sqrt{N_c}}{4\pi^2}\int\limits_{(m_1+m_2)^2}^\infty
\frac{ds\phi_V(s)}{(s-p^2-i0)}
(m_1+m_2)\left(\rho_+(s,m_1,m_2)+\frac{\rho_{k_\perp^2}}{\sqrt{s}+m_1+m_2}\right),  
\end{eqnarray}
with $\rho_+(s,m_1,m_2)$ and $\rho_{k_\perp^2}$ given in (\ref{rhoplus}) and (\ref{rhokperp2}). 

The light-cone representation (\ref{hp-lc}) leads to the following large-$m_Q$ behaviour of $H^{(Q,q)}_P$ for the heavy-light vector 
meson $\bar Qq$: 
\begin{eqnarray}
\label{Hp_HQ}
H_P^{(q)}\to \frac{f_V}{\bar \Lambda}\frac{m_Q^2}{m_Q^2-p^2},\qquad H_P^{(Q)}\to \frac{f_V m_Q}{m_Q^2-p^2}. 
\end{eqnarray}
For the $B$-decays of interest, we need the value of the form factors $H_{P,S}(p^2,q_1^2=0)$ at $p^2=M_B^2$,
which lies above the threshold $(m_c+m_q)^2$. The spectral representations for $H_{P,S}(p^2=M_B^2)$ develop the imaginary parts 
which occur due to the quark-antiquark intermediate states in the $p^2$-channel. It should be emphasized that no anomalous cuts 
emerge in the double spectral representation at $q_1^2\le 0$ \cite{lms_triangle}. In all cases considered in this paper,  
the value of $p^2=M_B^2$ lies far above the region of resonances which occur in the quark-antiquark channel.  
Far above the resonance region, local quark-hadron duality works well and the calculation of the imaginary part based on the 
quark diagrams is trustable. The imaginary part turns out to be orders of magnitude smaller than the real part of the form factor 
and for the practical purpose of the decay rate calculation may be safely neglected.

\section{\label{Sect5}Numerical results}
The derived spectral representations for the form factors allow one to obtain numerical predictions for the form factors of interest 
as soon as the parameters of the model -- the meson wave finctions and the quark masses -- are fixed. 
\subsection{Parameters of the model}
The wave function $\phi_i(s)$, $i=P,V$ can be written as  
\begin{eqnarray}
\label{phi}
\phi_i(s)=
\frac{\pi}{\sqrt2}
\frac{\sqrt{s^2-(m_1^2-m^2)^2}}{\sqrt{s-(m_1-m)^2}}
\frac{w_i(k^2)}{s^{3/4}}, \qquad 
k^2=\lambda(s,m_1^2,m^2)/4s, 
\end{eqnarray}
with $w_i(k^2)$ normalized as follows 
\begin{eqnarray}
\label{norm}
\int w^2_i(k^2)k^2 dk=1. 
\end{eqnarray}
The meson weak transition form factors from dispersion approach reproduce correctly the structure of the heavy-quark expansion 
in QCD for heavy-to-heavy and heavy-to-light meson transitions, as well as for the meson-photon transitions, if the 
radial wave functions $w(k^2)$ are localized in a region of the order of the confinement scale, $k^2\le\Lambda^2$ \cite{melikhov}. 

Following \cite{ms}, we make use of a simple gaussian parameterization of the radial wave function
\begin{equation}
\label{gauss}
w_i(k^2)\propto\exp(-k^2/2\beta_i^2),
\end{equation}
which satisfies the localization requirement for $\beta\simeq \Lambda_{QCD}$ and 
proved to provide a reliable picture of a large family of the transition form factors \cite{ms}. 

In \cite{ms} we fixed the parameters of quark model---constituent quark masses and the wave-function parameters $\beta_i$ 
of the Gaussian wave functions---by requiring that the dispersion approach reproduces  
(i) meson decay constants and (ii) some of the well-measured lattice QCD results for the form factors at large $q^2$.  
The analysis of \cite{ms} demonstrated that a simple Gaussian Ansatz for the radial wave functions allows one to reach this goal 
(to great extent due to the fact that the dispersion representations satisfy rigorous constrains from non-perturbative QCD in the 
heavy-quark limit). With these few model parameters, \cite{ms} gave predictions for a great number of weak-transition form factors 
in the full kinematical $q^2$-region of weak decays;  these results were shown to agree with the available results from lattice QCD 
and QCD sum rules within 10\% accuracy in the full $q^2$-region. We therefore assign a 10\%-uncertainty to our form-factor estimates in 
this work. 

We use here the same values of the constituent quark masses as obtained in \cite{ms}
\begin{equation}
\label{quark_masses}
\quad m_d=m_u=0.23 \;{\rm GeV}, \quad m_s=0.35 \;{\rm GeV},\quad m_c=1.45 \;{\rm GeV},\quad m_b=4.85 \;{\rm GeV}.
\end{equation} 
With the quark masses (\ref{quark_masses}) and the meson wave-function parameters $\beta$ quoted 
in Table \ref{table:parameters}, the decay constants from our dispersion 
approach reproduce the best-known decay constants of pseudoscalar and vector mesons also summarized in Table 
\ref{table:parameters}. 

\begin{table}[ht!]
\caption{\label{table:parameters}
Meson masses from \cite{pdg}, leptonic decay constants, and the corresponding wave-function parameters $\beta$ \cite{lmss}.}
\centering
\begin{tabular}{|c|c|c|c|c|c|c|}
\hline
            & $B$     &  $B_s$ &  $D^*$ &  $D_s^*$  &   $J/\psi$   \\
\hline
$M$, GeV    & 5.279   &  5.370 &  2.010 &   2.11    &  3.097  \\
$f$, MeV    
&  $192\pm 8$ \cite{lmsfB}      &   $226\pm 15$ \cite{lmsfB}   
&  $248\pm 2.5$ \cite{lmsfD*}   &   $311\pm 9$ \cite{lmsfD*}   
&  $405\pm 7$ \cite{fpsi_lat,pdg}  \\
\hline
$\beta$, GeV &   $0.565$    &   $0.62$  &   $0.48$    &   $0.54$     &  $0.68$   \\
\hline
\end{tabular}
\end{table}
\subsection{$B\to\gamma$ and $\gamma\to V$ form factors}
Before turning to the numerical estimates, let us emphasize that, as obvious from (\ref{Fa_HQ}),(\ref{Fv_HQ}), (\ref{Hp_HQ}), 
and (\ref{Hs_HQ}), the photon emission from the $V$-loop and from the $B$-loop have the same scaling behaviour 
in the heavy-quark limit. 
Therefore, {\it a priori} there is no valid reason to neglect the $V$-loop contributions. We shall see that indeed the photon emission 
from the vector-meson loop gives the contribution of a similar size as the photon emission from the $B$-meson loop.  
Our numerical estimates for the necessary form factors are summarized in Tables \ref{table:Fa+Fb} and \ref{table:Hp+Hs}.  

\begin{table}[htb]
\caption{\label{table:Fa+Fb} 
The form factors $F_A(M_V^2)$ and $F_V(M_V^2)$ describing the $B\to\gamma$ and $B_s\to\gamma$ transition
for $V=J/\psi,D^*_s,D^*$}
\centering
\begin{tabular}{|l|r|r|r||l|r|r|r|}
\hline
$B_s\to\gamma$ & $M_V^2=M_\psi^2$ & $M_V^2=M^2_{D_s^*}$   & $M_V^2=M^2_{D^*}$ & 
$B\to\gamma$   & $M_V^2=M_\psi^2$ & $M_V^2=M^2_{D_s^*}$   & $M_V^2=M^2_{D^*}$ \\
\hline
$F_V^{(b)}(M_V^2)$ &  $-0.060$   &  $-0.048$ &  $-0.046$ & $F_V^{(b)}(M_V^2)$   &  $-0.054$   &  $-0.044$ &  $-0.043$ \\
$F_V^{(s)}(M_V^2)$ &  $-0.410$   &  $-0.328$ &  $-0.322$ & $F_V^{(u,d)}(M_V^2)$ &  $-0.388$   &  $-0.316$ &  $-0.310$ \\
$F_A^{(b)}(M_V^2)$ &  $0.074$    &  $0.059$  &   $0.058$ & $F_A^{(b)}(M_V^2)$   &  $0.066$    &   $0.052$ &   $0.050$ \\
$F_A^{(s)}(M_V^2)$ &  $0.324$    &  $0.279$  &   $0.276$ & $F_A^{(u,d)}(M_V^2)$ &  $0.304$    &   $0.268$ &   $0.264$ \\
\hline
\end{tabular}
\end{table} 


\begin{table}[htb]
\caption{\label{table:Hp+Hs} 
The form factors $H_P(p^2)$ and $H_S(p^2)$, describing the $\gamma\to V$ transition ($V=J/\psi, D_s^*, D^*$) for $p^2=M_B^2$. 
The difference between the form factors at $p^2=M_B^2$ and $p^2=M_{B_s}^2$ is negligible and may be safely ignored. 
One finds ${\rm Im}\, H_{P,V}(M_B^2)\ll {\rm Re}\, H_{P,V}(M_B^2)$ and thus ${\rm Im}\, H_{P,V}(M_B^2)$ may be safely neglected for 
the decay-rate calculations.}
\centering
\begin{tabular}{|l|rr|rr|rr|}
\hline
         &  \multicolumn{2}{c|}{$\gamma\to J/\psi$} &  \multicolumn{2}{c|}{$\gamma\to D_s^*$ ($q=s$)} &  \multicolumn{2}{c|}{$\gamma\to D^*$ ($q=u,d$}  \\
\hline
                    & $p^2=M_B^2$ & $p^2=M_{B_s}^2$  & $p^2=M_B^2$  & $p^2=M_{B_s}^2$  &  $p^2=M_B^2$  & $p^2=M_{B_s}^2$  \\
\hline
$H_P^{(c)}(M_B^2)$   & $-0.196$ & $-0.183$ &  $-0.044$ &  $-0.042$ &  $-0.032$ &  $-0.030$ \\
$H_P^{(q)}(M_B^2)$   &  ---     &   ---    &  $-0.096$ &  $-0.092$ &  $-0.081$ &  $-0.078$ \\
$H_S^{(c)}(M_B^2)$   &  $0$     & $0$      &  $0.016$  &  $0.015$  &  $0.014$  &  $0.013$  \\
$H_S^{(q)}(M_B^2)$   &  ---     &   ---    &  $-0.007$ &  $-0.006$ &  $-0.002$ &  $-0.001$ \\
\hline
\end{tabular}
\end{table} 

\subsection{The decay rates}
\begin{center}
\begin{figure}[ht!]
\mbox{\epsfig{file=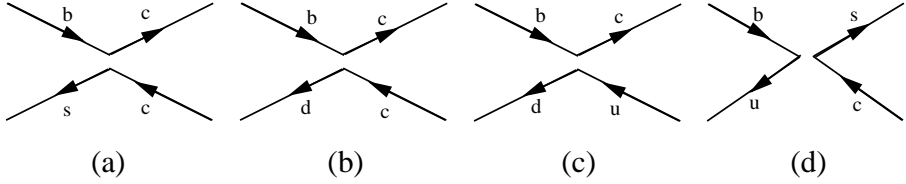,width=12cm}}
\caption{\label{fig:quark_diag}
Four-quark operators inducing the annihilation $B$-decays listed in Table \ref{table:wa-amplitudes}. 
(a) $\bar B^0_s\to J/\psi\gamma$; 
(b) $\bar B^0_d\to J/\psi\gamma$;  
(c) $\bar B^0_d\to D^{*0}\gamma$;  
(d) $B^{-}\to D_s^{*-}\gamma$. }
\end{figure}
\end{center}

We have now everything for the calculation of the amplitudes and the decay rates. 
We consider several annihilation-type $B$-decays which have the highest probabilities; the weak-annihilation 
quark diagrams which induce these decays are shown in Fig.~\ref{fig:quark_diag}. 

The corresponding $F_{PC}$ and $F_{PV}$ and the decay rates are summarized in Table \ref{table:wa-amplitudes}. To highlight the 
contribution to the amplitudes coming from the photon emission from the $V$-meson loop, we multiply it by a coefficient $r$ 
which is set to unity in the decay-rate calculations. 
Obviously, for some modes the photon emission from the vector-meson loop is comparable or even exceeds the photon emission from the $B$-meson loop and thus should be taken into account. 

\begin{table}[h!]
\caption{\label{table:wa-amplitudes} 
The amplitudes and the branching ratio for the annihilation-type decay of $B$ and $B_s$.}
\centering
\begin{tabular}{|c|c|c|c|c|}
\hline
Reaction               & CKM-factor            &    $F_{PC}$ [GeV]   &    $F_{PV}$ [GeV]&       Br       \\
\hline
$\bar B^0_s\to J/\psi\gamma$  &  $a_2 V_{cb}V_{cs}^*$  &   $0.036-0.052 r$   &       0.020     &   $1.43\cdot 10^{-7}\left(\frac{a_2}{0.15}\right)^2$  \\
$\bar B^0_d\to J/\psi\gamma$  &  $a_2 V_{cb}V_{cd}^*$  &   $0.035-0.050 r$   &       0.021     &   $7.54\cdot 10^{-9}\left(\frac{a_2}{0.15}\right)^2$  \\
$\bar B^0_d\to D^{*0}\gamma$  &  $a_2 V_{cb}V_{ud}^*$  &   $0.012-0.014 r$ &   $0.007+0.002 r$ &   $4.33\cdot 10^{-8}\left(\frac{a_2}{0.15}\right)^2$  \\
$B^{-}\to D_s^{*-}\gamma$  &  $a_1 V_{ub}V_{cs}^*$  &  $-0.025+0.001 r$ & $-0.014+0.002 r$     &  $1.68\cdot 10^{-7}\left(\frac{a_1}{1.02}\right)^2$  \\
\hline
\end{tabular}
\end{table} 
For the scale-dependent Wilson coefficients $C_i(\mu)$ and $a_{1,2}(\mu)$ at the renormalization scale $\mu\simeq 5$ GeV 
we use the following values \cite{heff}: $C_1=1.1$, $C_2=-0.241$, $a_1=C_1+C_2/N_c=1.02$, and $a_2=C_2+C_1/N_c=0.15$. 
Similar values are used for numerical etimates in \cite{ch2004}: e.g., for $B_{(s)}\to J/\psi\gamma$ decay, 
$a_2=0.15$ in our analysis corresponds to the effective Wilson coefficient $\bar a_q=0.163$.


\section{\label{Sect6}Discussion and Conclusions}
We have analyzed the annihilation-type radiative $B$ decays in the naive factorization approximation, taking into account 
both the photon 
emission from the $B$-meson loop and the vector-meson loop ($V$-loop). The latter contribution was not taken into account 
in all previous 
analyses and is therefore the novel feature of this paper. We have shown that in general, the photon emission from the $V$-loop 
is not 
suppressed compared to the photon emission from the $B$-loop and gives a comparable contribution. Our main results are as follows: 

\begin{itemize}
\item[(i)] 
We calculated the form factors $F_A$ and $F_V$ describing the photon emission from the $B$-loop, 
and the form factors $H_P$ and $H_S$ 
describing the photon emission from the $V$-loop making use of the relativistic dispersion approach 
based on the constituent quark picture. 
The form factors from this method satisfy all rigorous constrains from QCD in 
the heavy-quark limit for heavy-to-heavy, heavy-to-light, 
and heavy-meson-photon transition form factors. 
The numerical parameters of the model such as the effective constituent quark masses and 
the nonperturbative meson wave functions have been fixed in \cite{ms} by fitting to 
well-known leptonic decay constants of heavy mesons and 
a few well-measured form factors from lattice QCD. The predictions from the dispersion 
approach to the transition form factors 
have been then tested in many $B$ and $D$ decays and agree quite well with the available 
results from lattice QCD and QCD sum rules with 
the accuracy of a few percent \cite{m}. So, we assign a 10\% uncertainty in the decay rate 
related to the form-factor uncertainties. 

We emphasize that the photon emission from the $V$-loop has no parametric suppression compared 
to the photon emission from the $B$-loop  
and therefore cannot be neglected. Moreover, the numerical impact of the photon emission from 
the $V$-loop is substantial; for instance, 
in the case of the $\bar B_s\to J/\psi\gamma$ decay, taking into account the photon emission 
from both $s$- and $b$-quarks in the $B$-loop and the photon emission from the $V$-loop, 
leads to a strong 60\% suppression of the 
decay rate compared to the result based on merely the photon emission by the 
light quark of the $B$-meson. 

\item[(ii)] 
Making use of our results for the form factors and employing naive factorization for the 
complicated amplitudes of the 4-quark operators, 
we obtain predictions for the annihilation-type decays with the largest branching fractions: 
\begin{eqnarray}
\label{br1}
{\cal B}(\bar B^0_s\to J/\psi\gamma)    &=&1.43\cdot 10^{-7}\left(\frac{a_2}{0.15}\right)^2, \\
\label{br2}
{\cal B}(\bar B^0_d\to J/\psi\gamma)    &=&7.54\cdot 10^{-9}\left(\frac{a_2}{0.15}\right)^2, \\
\label{br3}
{\cal B}(\bar B^0_d\to D^{0*}\gamma)    &=&4.33\cdot 10^{-8}\left(\frac{a_2}{0.15}\right)^2,  \\
\label{br4}
{\cal B}(B^-\to \bar D_s^{*-}\gamma)    &=&1.68\cdot 10^{-7}\left(\frac{a_1}{1.02}\right)^2.
\end{eqnarray}
We would like to emphasize a relatively large branching ratio of the $\bar B^-\to \bar D_s^{*-}\gamma$ decay which 
makes this mode a prospective candidate for the experimental studies in the near future.

\item[(iii)] 
Uncertainties in our predictions listed above come from the two sources: 
(a) as just mentioned above, 
an approximate model for the form factors 
which yields an error in the decay rate at the level of 10-15\% and 
(b) naive factorization of the complicated 4-quark operators. 
The accuracy of the naive factorization for the decay rates may be probed to some extent by variations of the scale $\mu$ in 
the scale-dependent Wilson coefficients $C_i(\mu)$ (recall that the amplitudes $A_{\rm PV}$ and $A_{\rm PC}$ 
are scale-independent). Another way to access the size of the nonfactorizable corrections was indicated in \cite{grinstein}, 
where the nonfactorizable corrections in heavy-to-heavy radiative decays have been related to nonfactorizable corrections in 
the $B-\bar B$ oscillations. The latter have been found to be at the level of a few percent \cite{mn_oscillations}. 
On the basis of these arguments, one does not expect corrections to factorization larger than 5-10\%. 
We therefore assign here a 10\% uncertainty to the branching ratios related to nonfactorizable contributions. 

In view of this argument, huge negative corrections to factorization in $B^0_s\to J/\psi \gamma$ reported 
in \cite{ch2004}, which lead to a suppression of the decay rate by almost a factor 30, seem unrealistic. 
The correction to naive factorization in \cite{ch2004} has been calculated within the formalism of \cite{ch2004}. 
Ref.~\cite{ch2004} reported a strong cancellation between the factorizable contribution and the radiative 
correction calculated using QCD factorization \cite{bbns}. However, the authors of \cite{ch2004} have not taken into account 
several other contributions to the amplitude (e.g., the photon emission from the charm loop). 
Therefore, the huge reduction of the branching ratio reported in \cite{ch2004} does not seem to us trustable and the analysis 
of nonfactorizable effects should be revised. Ref.~\cite{ch2004} reports also 
the branching ratios based on factorization approximation; however the factorization results of \cite{ch2004} 
neglect several effects 
(photon emission from the heavy quark of the $B$-loop and photon emission from the $V$-loop) which lead to a 
visible suppression of the branching ratio. So we do not confirm also the factorization results of \cite{ch2004}. 

Ref.~\cite{ch2006} reported another estimate for $\bar B_s\to J/\psi\gamma$ based on naive factorization, 
neglecting the photon emission from the $V$-loop.\footnote{Ref.~\cite{ch2006} contains an erroneous statement 
that the photon emission from the $V$-loop vanishes for the equal quark masses in the vector meson. 
The cancellation between the photon emission from the quark and from the antiquark of the $V$-meson occurs 
for the form factor $H_S$, whereas for the form factor $H_P$ the amplitudes of these two subprocesses add to each other.} 
However, the huge form factors $F_{A,V}$ reported in \cite{ch2006} clearly contradict the results from the large-energy 
effective theory \cite{korch,mk} and therefore the results of \cite{ch2006} cannot be trusted. 
\end{itemize}
In conclusion, we believe that in comparison with the existing estimates \cite{ch2004,ch2006} credit should be given to our results: 
First, we take into account those contributions which have been neglected in \cite{ch2004,ch2006} 
but which are shown to give sizeable contributions to the amplitude. Second, our calculation of the form factors is based on 
a more detailed model for the $B$-meson structure than the models employed in \cite{ch2004,ch2006}.
Taking into account the uncertainties mentioned above (the scale in the Wilson coefficients, 
making use of the factorization approximation for the weak-annihilation amplitude, uncertainties in the form factors), 
we estimate the accuracy of our theoretical predictions for the branching ratios (\ref{br1})--(\ref{br4}) to be at 
the level of 20\%.

\acknowledgments 
D.~M. has pleasure to thank Berthold~Stech for stimulating discussions on the subject of this paper. 
The work was supported by a Grant for Leading Scientific Schools 3042.2014.2 (Russia). 
N.~N. was supported in part by Russian Ministry of Education and Science under grant 14.610.21.0002.

\end{document}